\documentclass[aps,prl,twocolumn,groupedaddress,10pt]{revtex4-1}

\usepackage{graphicx}
\usepackage[latin1]{inputenc}
\usepackage{grffile}

\usepackage{amsfonts}
\usepackage{amsmath}
\usepackage{amssymb}
\usepackage{color}
\usepackage[normalem]{ulem}
\usepackage{todonotes}
\usepackage[bookmarks, 
            breaklinks, 
            pdfstartview=FitH, 
            pdftitle={Integrated Mach-Zehnder interferometer for Bose-Einstein condensates},
            pdfauthor={T.~Berrada et al.},
            pdfkeywords={}
            ]{hyperref}

\renewcommand{\figurename}{Figure}

\begin{document}

\title{Integrated Mach-Zehnder interferometer for Bose-Einstein condensates}

\author{T.~Berrada}
\author{S.~van~Frank}
\author{R.~Bücker}
\author{T.~Schumm}
\author{J.-F.~Schaff}
\email[]{jschaff@ati.ac.at}
\author{J.~Schmiedmayer}
\affiliation{Vienna Center for Quantum Science and Technology, Atominstitut, TU Wien, Stadionallee 2, 1020 Vienna, Austria}

\date{\today}

\begin{abstract}
\textbf{Particle-wave duality enables the construction of interferometers for matter waves, which complement optical interferometers in precision measurement devices. This requires the development of atom-optics analogs to beam splitters, phase shifters, and recombiners. Integrating these elements into a single device has been a long-standing goal. Here we demonstrate a full Mach-Zehnder sequence with trapped Bose-Einstein condensates (BECs) confined on an atom chip. Particle interactions in our BEC matter waves lead to a non-linearity, absent in photon optics. We exploit it to generate a non-classical state having reduced number fluctuations inside the interferometer. Making use of spatially separated wave packets, a controlled phase shift is applied and read out by a non-adiabatic matter-wave recombiner. We demonstrate coherence times a factor of three beyond what is expected for coherent states, highlighting the potential of entanglement as a resource for metrology. Our results pave the way for integrated quantum-enhanced matter-wave sensors.}
\end{abstract}

\maketitle

Atom interferometers~\cite{Cronin2009} are powerful tools for a wide variety of measurements. They have been used to measure atomic properties~\cite{Ekstrom1995}, to study quantum degenerate systems~\cite{Hofferberth2008,Gring2012}, for precision measurements~\cite{Weiss1993}, and are very sensitive probes for inertial effects~\cite{Clauser1988}, as beautifully demonstrated for example in gravimeters~\cite{Peters1999} and gyroscopes~\cite{Gustavson2000}.

Whereas interferometers with atoms propagating in free fall are ideally suited for inertial and precision measurements ~\cite{Weiss1993,Clauser1988,Peters1999,Gustavson2000,Fixler2007}, interferometers where the atoms are held in tight traps or guides are better for measuring weak localized interactions, as for example proposed to probe the Casimir-Polder force~\cite{Carusotto2005}.

In most atom interferometers to date, the matter waves propagate freely and the two interfering arms correspond to wave packets having different momenta~\cite{Cronin2009}. The beam splitters creating the superposition either couple different internal states, like in the Raman beam splitter, or leave the matter wave in the same atomic state, like in Bragg scattering or diffraction from a material grating.

Bose-Einstein condensates (BECs) are promising candidates for atom interferometry owing to their macroscopic coherence properties~\cite{Andrews1997,Hagley1999, Bloch2000, Perrin2010}. Following the first observation of BEC interference in 1997~\cite{Andrews1997}, various building blocks of BEC interferometers have been realized individually. Interference experiments with BECs were performed using Bragg beams in ballistic expansion~\cite{Hagley1999}, with atoms freely propagating in a guide~\cite{Wang2005, McDonald2013}, and very recently under microgravity~\cite{Muntinga2013}. Splitting a single trapped BEC into two separated clouds in a double well, interference was observed using ballistic expansion after switching off the trapping potential~\cite{Shin2004, Schumm2005, Jo2007b, Jo2007, Baumgartner2010}. Realizing a full Mach-Zehnder interferometer with spatially separated modes of trapped BECs has been a long-standing goal~\cite{Pezze2005, Grond2010, Grond2011}, especially in chip traps, where the tight confinement allows local probing. 

A fundamental difference between photon and matter-wave optics is the presence of atom-atom interactions. In dense, confined geometries they lead to dephasing (phase diffusion) which ultimately limits the coherence time of the interferometers~\cite{Lewenstein1996, Castin1997, Javanainen1997, Baumgartner2010}. Phase diffusion can be reduced by controlling the interactions with a Feshbach resonance. Coherence times up to $10$\,s were recently demonstrated in Bloch oscillation experiments~\cite{Fattori2008, Gustavsson2008}.

Interactions can also be used to generate non-classical correlations between the two paths of the interferometer, which can be used either to improve the sensitivity~\cite{Kitagawa1993} or to reduce the effect of phase diffusion~\cite{Jo2007b}. Generation of entanglement was recently reported for BECs in multiple-well potentials~\cite{Esteve2008} and between two or three internal states of spinor BECs~\cite{Gross2010, Riedel2010, Gross2011, Lucke2011, Hamley2012}. In Ref.~\cite{Gross2010} entanglement was used in an internal state BEC interferometer to improve its sensitivity below the standard quantum limit $\delta \phi_\text{SQL} = 1/\sqrt{N}$, where $N$ is the number of atoms involved in the measurement.

Here we demonstrate a full Mach-Zehnder interferometer for Bose-Einstein condensates integrated on an atom chip. Our interferometric scheme relies on the coherent splitting and recombination of a Bose-Einstein condensate in a tunable magnetic double-well potential, where the matter wave is confined at all times. Thanks to a spatial separation of $\sim 2\,\mu$m between the two wave packets, our geometry is sensitive to accelerations and rotations. By tilting the double well out of the horizontal plane for a variable time $t_{\phi}$, we apply an energy difference $\epsilon$ and thereby imprint a controlled relative phase between the interferometer arms. A non-adiabatic recombiner translates the relative phase into an atom number difference, which is directly read out using a highly sensitive time-of-flight fluorescence detector~\cite{Bucker2009}. We show that the use of a non-classical state with reduced number fluctuations inside the interferometer helps increasing the interrogation time, which is still limited by interaction-induced dephasing. This represents an unambiguous demonstration of the interplay between number fluctuations and phase diffusion in the case of spatially separated modes of Bose-Einstein condensates.

\section{Results}

\subsection{Principle and interferometric fringes}

\begin{figure*}[ht]
\includegraphics[width=\linewidth]{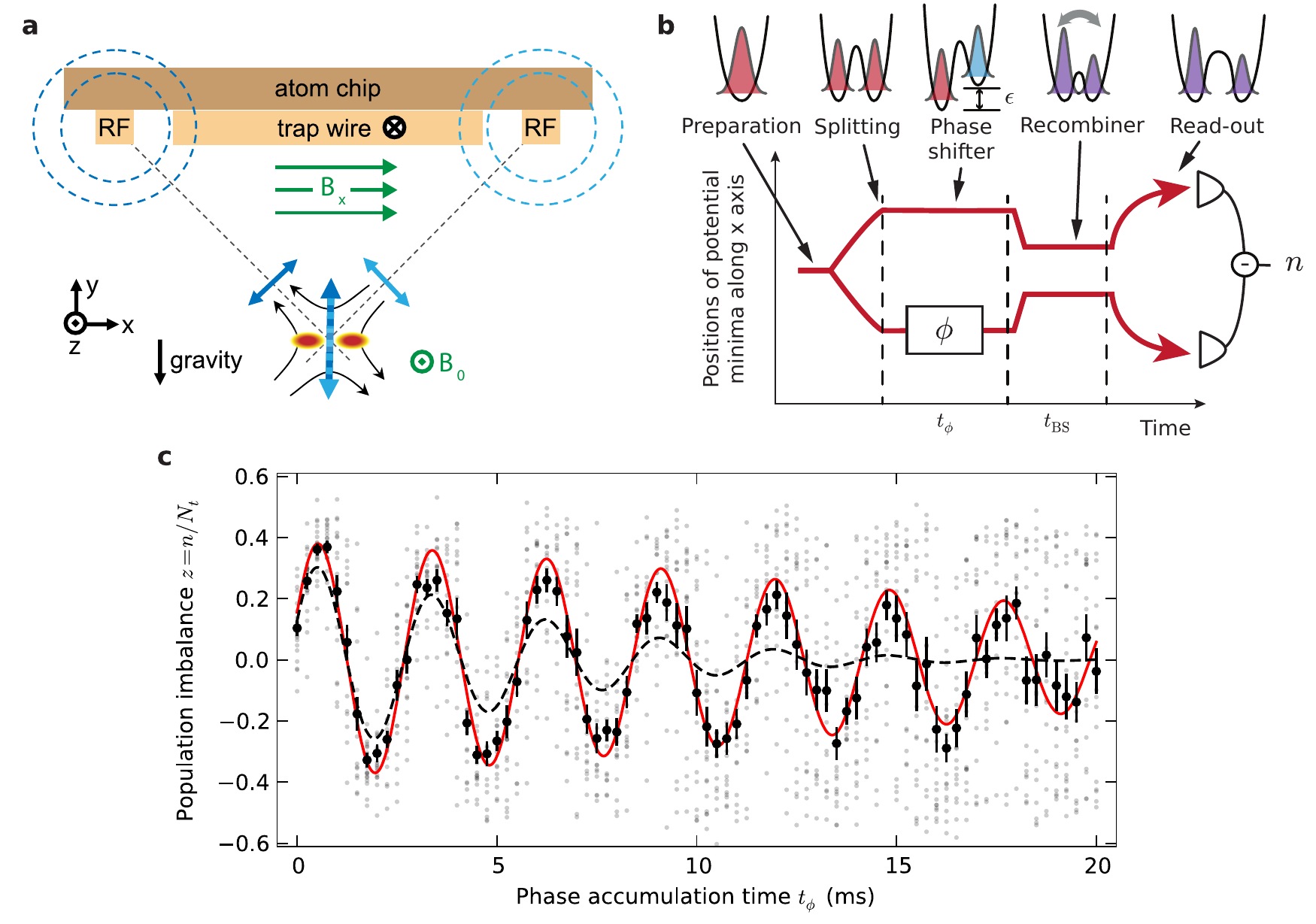}
\caption{\textbf{The Mach-Zehnder interferometer and its output signal.} \textbf{a,} Schematic of the atom chip: a DC current of $1$\,A is sent through the trap wire, which, together with a uniform  external ``bias'' field $B_x = 29.5$\,G, creates an elongated quadrupole trap $60\,\mu$m below the chip. The longitudinal confinement is realized by two wires parallel to the $x$ axis (not shown here). An external field $B_0$ completes the Ioffe-Pritchard configuration. Rf-currents with a relative phase of $\pi$ are applied on the dressing wires to perform the splitting along $x$. \textbf{b,} The condensate is coherently split by transforming a single trap into a double-well potential; a relative phase between the two arms is imprinted by tilting the double well during a time $t_{\phi}$; the spacing between the two wells is then abruptly reduced and the potential barrier acts as a beam splitter for both wave packets, transforming the relative phase into a population imbalance. After the recombination time $t_\text{BS}$, the atom clouds are separated and the particle number in each well is read out by fluorescence imaging. \textbf{c,} The normalized population difference between the two wells $z = n/N_\text{t}$ is measured as a function of $t_{\phi}$. It exhibits interference fringes and a damping due to phase diffusion. Grey dots: imbalance of individual experimental realizations; black dots: ensemble average $\langle z \rangle$; red curve: theoretical prediction taking into account phase diffusion; dashed black line: expected signal for a classical coherent state for which $\xi_\text{N} = 1$. The error bars indicate $\pm 1$ standard error of the mean. \label{Fig:Schematics}}
\end{figure*}

The atom chip used to manipulate the atomic wave packets is depicted in Fig.~\ref{Fig:Schematics}a. The interferometer sequence is illustrated in Fig.~\ref{Fig:Schematics}b, and Fig.~\ref{Fig:Schematics}c displays the resulting oscillation of the mean of the normalized population difference $ z  \equiv  (N_\text{L} - N_\text{R})/N_\text{t} $ as a function of the phase accumulation time $t_{\phi}$, where $N_\text{L}$ and $N_\text{R}$ are the populations of the left and right wells respectively and $N_\text{t} \equiv N_\text{L} + N_\text{R}$.

In the following we will discuss each step of the sequence in more detail. We first focus on the BEC splitter and find that the state generated inside the interferometer is characterized by a spin-squeezing factor $\xi_\text{S}^2 = -7.8\pm0.8$\,dB~\cite{Kitagawa1993}, implying that it is entangled~\cite{Sorensen2001}. We then turn to the phase accumulation stage and analyse the phase diffusion process. We find a threefold extension of the coherence time using a number-squeezed state inside the interferometer as compared to a coherent state. Finally, we introduce the non-adiabatic recombiner and discuss the obtained sensitivity and possible improvements.

\subsection{Matter-wave source}

The experimental sequence starts with a degenerate gas of $^{87}$Rb atoms in the $F = 1$ hyperfine state in an elongated magnetic trap created by our atom chip set-up (cf.\ Methods)~\cite{Trinker2008a}. Using radio-frequency (rf) evaporative cooling, we prepare samples containing $N \equiv \langle N_\text{t} \rangle = 1,200$ atoms at a temperature of $T \simeq 20$\,nK $=h/k_B\times420$\,Hz and a chemical potential of $\mu/h \simeq 500$\,Hz, $k_\text{B}$ being Boltzmann's constant and $h$ Planck's constant. The brackets denote ensemble averaging. Given the elongated trap geometry, the cloud is in the one-dimensional quasi-condensate regime~\cite{Petrov2000} (cf.\ Supplementary Note 5).

\subsection{Splitting}

\begin{figure}[ht]
\includegraphics[width=\linewidth]{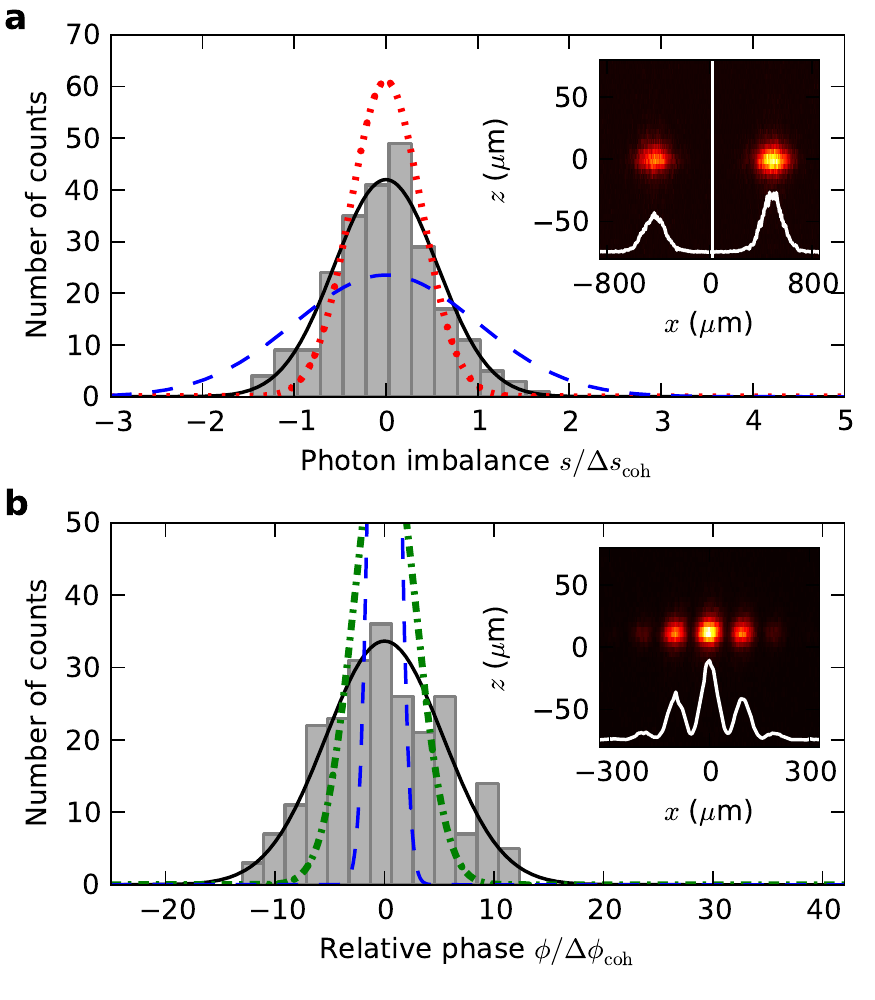}
\caption{\textbf{Number and phase distributions after splitting.} \textbf{a,} Histogram of the difference between the fluorescence signals of the left and right clouds $s = s_\text{L} - s_\text{R}$, in units of the standard deviation expected for a coherent state ${\Delta s}_\text{coh}$. The curves indicate a normal distribution corresponding to the measured number squeezing factor $\xi_\text{N} = 0.41 \pm 0.04$ (solid black); the distribution expected in the limit $\xi_\text{N} = 0$, where only detection noise is responsible for fluctuations (dotted red); and the distribution expected for a coherent state in the absence of detection noise (dashed blue). The inset shows a typical fluorescence picture and the regions used to define $s_\text{L}$ and $s_\text{R}$. \textbf{b,} Histogram of the measured relative phases $\phi$ in units of the circular standard deviation of a coherent state ${\Delta \phi}_\text{coh}$. The curves indicate a normal distribution with the measured standard deviation $\Delta \phi = 5.4 \pm 0.5 \times {\Delta \phi}_\text{coh}$ (solid black); and the distributions expected for a coherent state in the absence (dashed blue) and in the presence (dash-dot green) of detection noise. The inset shows a typical matter-wave interference pattern from which the phase is extracted.\label{Fig:Fluctuations}}
\end{figure}

The entire interferometer sequence relies on rf dressing~\cite{Lesanovsky2006, Hofferberth2006a} to dynamically turn the static magnetic trap into a double-well potential, whose spacing, barrier height, and tilt can be controlled by changing the amplitude and orientation of the rf field. Splitting is performed by increasing the rf amplitude in the dressing wires linearly within $5$\,ms to transform the single trap into a symmetric double-well potential~\cite{Schumm2005}. Along the horizontal splitting direction, the potential has two minima separated by $2\,\mu$m, the barrier height is $h\times3.7$\,kHz and tunnelling is negligible (cf.\ Methods).

To characterize the quantum state obtained after splitting, we measure the distributions of the two observables atom number difference $n \equiv N_\text{L} - N_\text{R}$ and relative phase $\phi \equiv \phi_\text{R} - \phi_\text{L}$. The interferometric sequence can be interrupted at any time to measure these quantities using destructive time-of-flight imaging (see Methods). The relative phase obtained from the Fourier transform of the interference fringes is used to monitor the system during the sequence and must not be confused with the one measured after recombination (measuring the wells populations). Each experimental run gives access to either one or the other observable. We repeat the measurement $\sim 200$ times to obtain good estimates of both the mean and the variance of $n$ and $\phi$.

For non-interacting particles, the atom number difference $n$ is expected to follow a binomial distribution with mean $\langle n \rangle = 0$ and standard deviation $\Delta n = \sqrt{N}$ after splitting. As repulsive interactions energetically favour equal populations, reduced fluctuations are expected~\cite{Leggett1998, Esteve2008, Maussang2010}. Figure \ref{Fig:Fluctuations}a shows the distribution of detected photon number difference $s \equiv s_\text{L} - s_\text{R}$ between the left and right clouds (our fluorescence detector registers on average $16$ photons per atom, see Methods). Indeed, we observe a suppression of fluctuations by a factor of two compared to shot noise. After correction of the detection noise~\cite{Bucker2009, Bucker2011} we deduce a number squeezing factor $\xi_\text{N} \equiv \Delta n / \sqrt{N} = 0.41 \pm 0.04$ ($\xi_\text{N}^2 = -7.8\pm0.8$\,dB). This shows that the splitting process yields a number-squeezed state with reduced fluctuations along one quadrature compared to a classical coherent state.

We now turn to the phase distribution shown in Fig.~\ref{Fig:Fluctuations}b. Since the two variables $n$ and $\phi$ can be seen as quantum mechanically conjugate ($[\phi, n] = 2 i$)~\cite{Kitagawa1993} the Heisenberg uncertainty relation imposes $\Delta \phi \geq 1/\Delta n = 0.074$\,rad. We observe a Gaussian-shaped phase distribution with vanishing average and circular standard deviation~\cite{Fisher1995} $\Delta \phi = 0.16 \pm 0.01$ rad ($\Delta n \, \Delta \phi = 2.3$), twice as large as expected from the Heisenberg lower bound. Note that contrary to the number distribution data, the phase distribution data were not corrected for imaging noise. We conjecture that this broadening is due to finite initial temperature \cite{Pitaevskii2001, Gati2006} and detection noise.

Altogether, the output state of the beam splitter exhibits reduced number fluctuations and high coherence properties, characterized by the coherence factor $\langle \cos \phi \rangle = 0.987 \pm 0.001$. It has been shown~\cite{Wineland1994} that using such a state in an alternative interferometric scheme can allow for a sensitivity better than the standard quantum limit. The improvement in sensitivity is measured by the spin squeezing factor~\cite{Kitagawa1993} $\xi_\text{S} \equiv \xi_\text{N}/\langle \cos \phi \rangle = 0.41 \pm 0.04$ and corresponds to a gain of $7.8\pm 1$\,dB over the standard quantum limit. Note that the numerical value of $\xi_\text{S}$ is here identical to that of $\xi_\text{N}$ thanks to the large coherence factor of almost unity. Their statistical uncertainties also coincide because that of the coherence factor is negligible compared to that of $\xi_\text{N}$. Since spin squeezing, defined as $\xi_\text{S} < 1$ is an entanglement witness~\cite{Sorensen2001}, this also demonstrates that our state is non-separable after splitting. The measured fluctuations and coherence imply that our ensemble contains 150 entangled particles in the sense of Refs.~\cite{Sorensen2001b,Hyllus2012} and we can exclude entanglement of less than 67 atoms with 90\,\% probability.

\subsection{Phase shifter}

\begin{figure*}[ht]
\centering
\includegraphics[width=\textwidth]{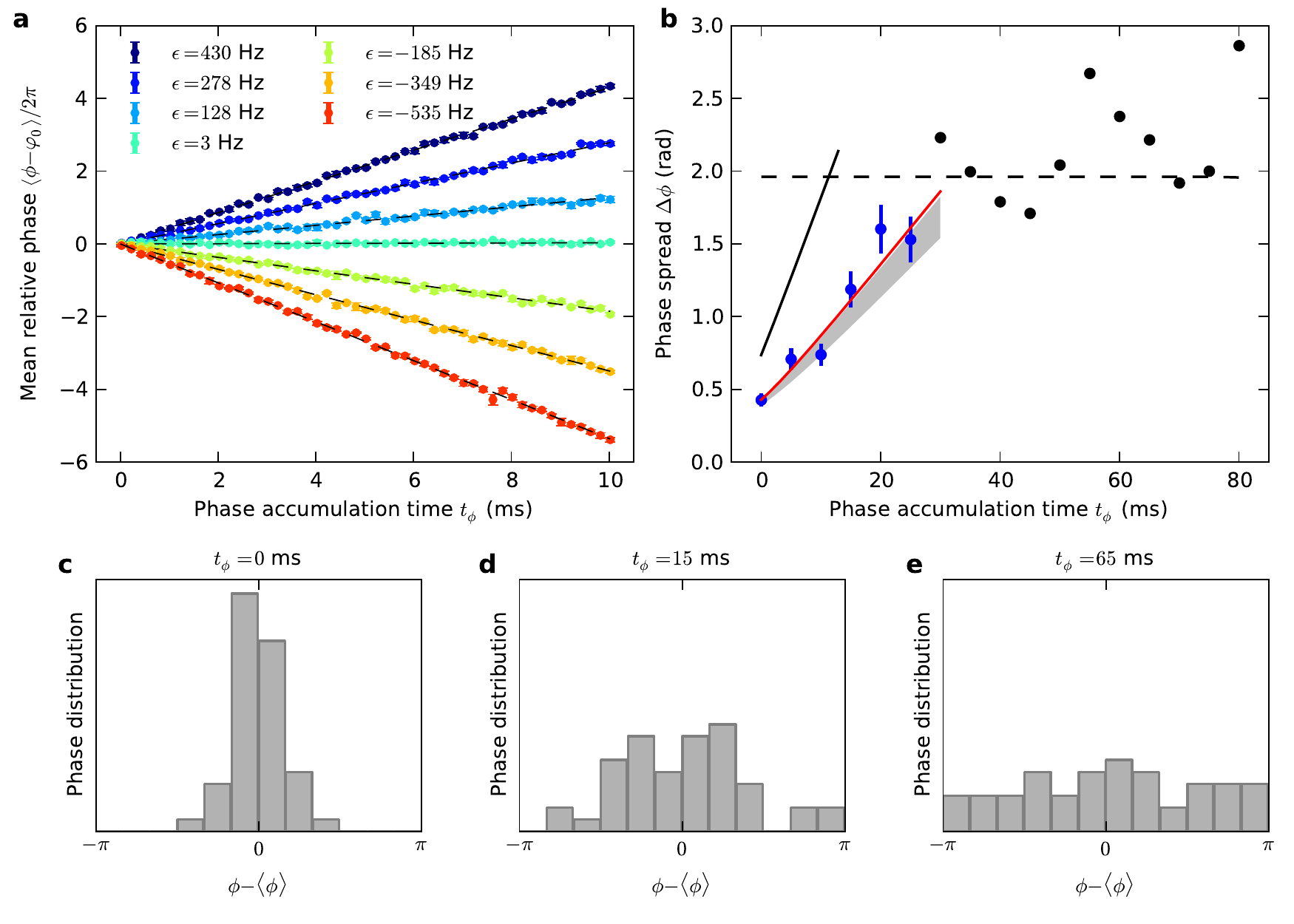}
\caption{\textbf{Evolution of the relative phase and its fluctuations.} All data on this figure have been obtained by time-of-flight recombination. \textbf{a,} Linear evolution of the phase for various energy differences $\epsilon$ induced by tuning the angle between the two wells. The values of $\epsilon$ shown in the legend are obtained from linear fits to the data (black dashed lines). For all fits, the standard error of the slope is below $3$\,Hz. The orange points correspond to the rate $\epsilon/h = -349 \pm 2$\,Hz used to record the fringes of Fig.~\ref{Fig:Schematics}c. \textbf{b,} Evolution of the circular standard deviation of the phase $\Delta \phi$ corresponding to the orange curve of panel a. It exhibits phase diffusion: at short times $t_\phi \lesssim 30$\,ms the mean phase is well defined (blue points); on the contrary, for longer times the phase distribution cannot be distinguished from a random distribution (black points, see Methods). The red line is a fit to the blue points with the model of Eq.~\eqref{Eq:PhaseSpread}. Shaded area: theoretical prediction without free parameter, taking into account the measured number squeezing (see text). Black line: expected behaviour if the initial state were classical (i.e.\ not number-squeezed). Note that at $t_{\phi} = 0$ phase diffusion has already started (see Eq.~\eqref{Eq:PhaseSpread}). \textbf{c,d,e,} Measured phase distributions 
for three values of $t_\phi$ indicated by the dashed lines. The error bars of a indicate $\pm 1$ circular standard deviation divided by $\sqrt{k_i}$, and those of b correspond to $\pm 1$ circular standard deviation divided by $\sqrt{2(k_i-1)}$, $k_i$ being the number of repetition used for each point $i$.
\label{Fig:PhaseShifter}}
\end{figure*}

We now discuss the phase accumulation stage. After splitting, the coherent superposition of left and right modes has on average $\langle  \phi \rangle = 0$ and $\langle n \rangle = 0$. We induce a deterministic shift of the relative phase by tilting the double-well trap out of the horizontal plane while keeping the well spacing constant. The tilt is performed within $3$\,ms by modifying the relative current in two rf wires on our atom chip. Given the inter-well distance of $2\,\mu$m and the angles of $\sim 10^\circ$, this corresponds to velocities at least 30 times smaller than the widths of the velocity distributions of the individual BECs, ensuring that the transfer is sufficiently adiabatic to prevent radial dipole oscillations of the condensates. The tilt is reversed before the recombining sequence. The energy difference $\epsilon$ which arises from the tilt (see Supplementary Note 6), triggers a linear evolution of the relative phase given by
\begin{equation}
\phi(t_\phi) = \varphi_\text{0} + \epsilon \, t_\phi/\hbar .
\label{Eq:PhaseShift}
\end{equation}
In Fig.~\ref{Fig:PhaseShifter}a, the phase evolution is measured for 7 different tilt angles ranging approximately from $-9^\circ$ to $+12^\circ$. For these measurements the phase is obtained directly from interference patterns measured after time-of-flight detection (cf.\ inset of Fig.~\ref{Fig:Fluctuations}b). Altogether this phase accumulation stage acts as a tunable phase shifter for the atomic interferometer. For the fringes presented in Fig.~\ref{Fig:Schematics}c the angle was fixed to yield the rate $\epsilon/h = -349 \pm 2$\,Hz and $t_\phi$ was varied to tune the phase.

\subsection{Phase diffusion}

In Fig.~\ref{Fig:PhaseShifter}b--e we observe that the linear evolution of the mean phase discussed above is accompanied by a broadening of the phase distribution, measured by the circular standard deviation of the phase $\Delta \phi$~\cite{Fisher1995}. This phenomenon illustrates phase diffusion~\cite{Lewenstein1996}. It corresponds to a dephasing arising from the interactions between the atoms: since the chemical potential of each condensate depends on its atom number, relative number fluctuations translate into fluctuations of the phase accumulation rate (cf.\ Supplementary Note 1 for the influence of a residual coupling between the condensates). This leads to an increase of the phase variance given by~\cite{Castin1997, Javanainen1997, Leggett1998}
\begin{equation}
\Delta \phi^2 (t_\phi) = \Delta \phi_0^2 + R^2 (t_\phi - t_\text{i})^2 ,
\label{Eq:PhaseSpread}
\end{equation}
where the phase diffusion rate $R$ is
\begin{equation}
R = \frac{\xi_\text{N} \sqrt{N}}{\hbar} \left.\frac{\partial \mu}{\partial \mathcal{N}}\right|_{\mathcal{N}=N/2} , \label{eq:rate}
\end{equation}
$\mu(\mathcal{N})$ being the chemical potential of one BEC with $\mathcal{N}$ atoms. The term $\Delta n = \xi_\text{N} \sqrt{N}$ corresponds to the standard deviation of the atom number difference $n$. One sees from Eqs.~\eqref{eq:rate} and \eqref{Eq:PhaseSpread} that number squeezing, defined as $\xi_\text{N} < 1$, reduces phase diffusion and allows longer interrogation times. From a fit (red line in Fig.~\ref{Fig:PhaseShifter}b) we obtain the phase diffusion rate $R = 51 \pm 4\,\text{mrad} \cdot \text{ms}^{-1}$ and an initial phase spread $\Delta \phi_0 = 300 \pm 90$\,mrad. The constant $t_\text{i} = -6$\,ms accounts for the $3$\,ms used before and after the phase accumulation stage to incline and level the double well.

The measured rate is in fair agreement with the expected value of $R = 46 \pm 4\,\text{mrad} \cdot \text{ms}^{-1}$ evaluated from Eq.~\eqref{eq:rate}. Because the splitting process is not adiabatic with respect to the axial direction, a slow axial quadrupole mode is excited and the condensates are not strictly at equilibrium during phase accumulation. Consequently, the term $\partial \mu/\partial \mathcal{N}$ in Eq.~\eqref{eq:rate} had to be slightly modified to evaluate Eq.~\eqref{eq:rate} (cf.\ Supplementary Note 2). This theoretical prediction with no free parameters is represented as a grey shaded area in Fig.~\ref{Fig:PhaseShifter}b.
To further confirm our understanding of phase diffusion, we directly measured $\partial \mu/\partial \mathcal{N}$ by splitting the initial BEC with an angle leading to a controlled population imbalance, levelling the double well back to horizontal ($\epsilon = 0$), and monitoring the evolution of the phase due to a difference of chemical potential. Together with the squeezing factor and the atom number measured, this gives a third value of $R = 57 \pm 5\,\text{mrad} \cdot \text{ms}^{-1}$ also in agreement with the two others.
For comparison, we evaluate the hypothetical value $R = 112\,\text{mrad} \cdot \text{ms}^{-1}$ in the absence of number squeezing ($\xi_\text{N} = 1$), indicated as a solid black line in Fig.~\ref{Fig:PhaseShifter}b. The difference with our measurements underlines that the interrogation time of our interferometer is extended by the fact that the initial state is number squeezed. Enhanced coherence times have been reported in Ref.~\cite{Jo2007b} and attributed to number squeezing. Our measurement of both number and phase fluctuations corroborates the direct link between phase diffusion and number squeezing.

\subsection{Recombiner}

\begin{figure*}[ht]
\includegraphics[width=\linewidth]{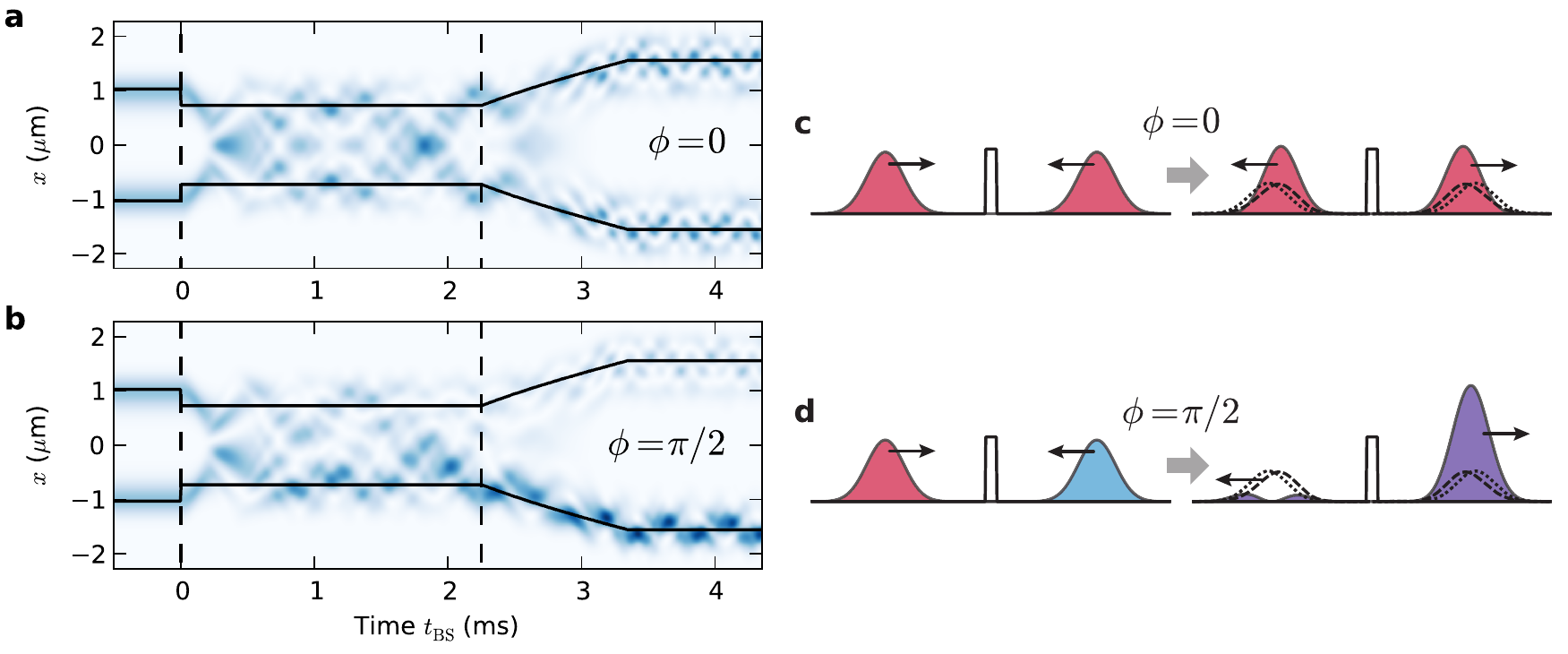}
\caption{\textbf{Principle of the BEC recombiner.} \textbf{a,b,} Numerical simulation of the evolution of the transverse in situ density when the double-well spacing is abruptly decreased ($0 < t_\text{BS} < 2.25$\,ms) and increased again ($t_\text{BS} > 2.25$\,ms) for two different initial relative phases: $\phi = 0$ (a), and $\phi = \pi/2$ (b). Black lines: positions of the two potential minima. \textbf{c,d,} This phase sensitive recombination can be understood as the propagation and interference of two wave packets launched onto a semi-reflective tunnel barrier. The final wave packets (filled curves with outgoing arrows) are the result of the constructive or destructive interference between the reflected (dashed lines) and transmitted (dotted lines) components arising from the incoming wave packets. \label{Fig:BeamSplitter}}
\end{figure*}

After the condensates have accumulated a relative phase, the last step consists in recombining the left and right modes in a way that converts the phase into an atom number difference. The recombiner plays a role similar to that of the output beam slitter of an optical Mach-Zehnder interferometer. In a two-mode picture, this operation corresponds to a $\pi/2$ pulse. In Ref.~\cite{Jo2007} the phase-dependent heating caused by the merging of two condensates was used  to infer the relative phase. It has also been proposed~\cite{Pezze2005, Grond2011} to use a quarter of a Josephson oscillation for this purpose. Nevertheless one expects the contrast of a Josephson recombiner to be fundamentally limited by the onset of self-trapping when the interaction energy is not negligible compared to the tunnelling energy~\cite{Raghavan1999}. Here we introduce a different method which makes use of a non-adiabatic transformation of the potential. We abruptly decrease the rf-dressing intensity to reduce the well spacing to 1.5\,$\mu$m and the barrier height to about $h \times$\,1\,kHz (see Fig.~\ref{Fig:BeamSplitter}a--b). We estimate the energy gap between the two lower-lying single-particle eigenstates of the double-well potential to be approximately $h \times 80$\,Hz. The clouds are accelerated towards the barrier and, after an adjustable time $t_\text{BS}$, the barrier is raised again to separate the atoms for counting. Starting with a state having a phase close to $\pi/2$, we adjusted $t_\text{BS}$ in order to maximize the final population imbalance. This yielded $t_\text{BS} = 2.25$\,ms. Repeating this operation for different initial phases, we observe the sine-like dependence of the population imbalance shown in Fig.~\ref{Fig:Schematics}c.

This effect can be understood as the splitting of the wave packets on a semi-reflective potential barrier and the interferences between them (see Fig.~\ref{Fig:BeamSplitter}c--d). Indeed, in the absence of interactions and assuming two symmetrically incoming wave packets with relative phase $\phi$, the linearity of the Schrödinger equation and symmetry of the potential assures that the population difference be given at any time by
\begin{equation}
n(t_{BS}) = N_\text{t} C(t_\text{BS}) \sin \phi .
\label{Eq:contrast}
\end{equation}
The contrast of the read-out $-1 \leq C(t_\text{BS}) \leq 1$ depends on the time spent in the coupled double well and is a function of the overlap of the transmitted and reflected modes on each side of the barrier at time $t_\text{BS}$ (cf.\ Supplementary Note 3). In a realistic double-well potential, the wave packets undergo a complicated dynamics involving multiple transmissions and reflections, resembling Bragg beams splitters (see Fig.~\ref{Fig:BeamSplitter}a--b). Simulations calibrated with the measured trap parameters (see  Methods and Supplementary Note 4) predict a maximum contrast of $80\,\%$, limited by the imperfect mode-matching in a double well with a finite barrier thickness. Taking into account interactions, no simple analytical model could be found. However, for weak interactions like in our case, numerical simulations yield a sine-like dependence of $n(\phi)$ and up to $70\,\%$ contrast. Experimentally, we could achieve a contrast of $42\,\%$ (see below). Possible reasons for this reduction are effects beyond mean-field and the coupling to the other spatial directions, both of which are not captured by our simulations.

\subsection{Interferometric signal}

\begin{figure}[ht]
\includegraphics[width=\linewidth]{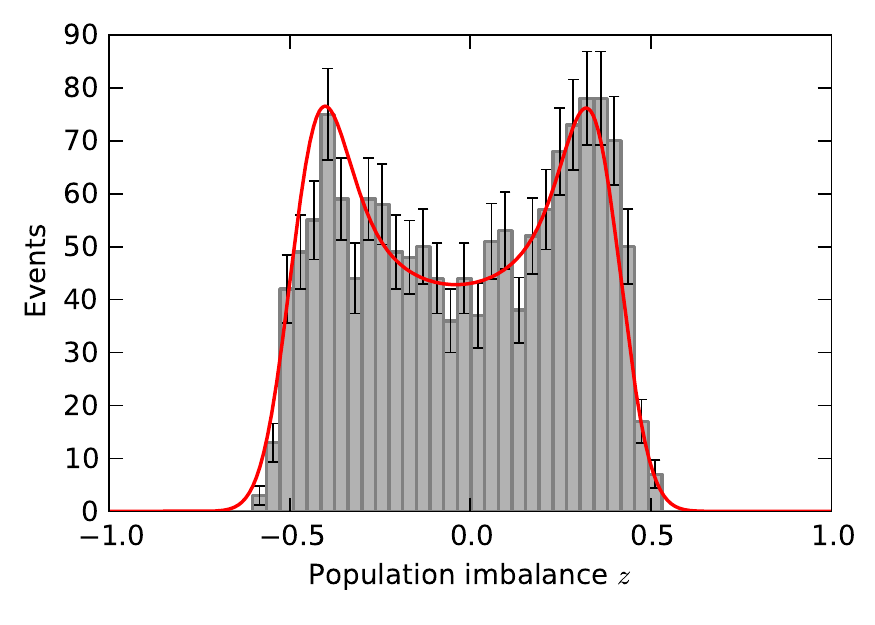}
\caption{\textbf{Distribution of the population imbalance.} Histogram of the single-run population imbalance $z$ (gray dots of Fig.~\ref{Fig:Schematics}c). It exhibits the characteristic double-structure expected from sampling the sine of a uniformly distributed phase. The contrast and the noise of the recombiner are estimated from convolving the probability distribution function given in Eq.~\eqref{Eq:pdfZ} with a Gaussian noise (red curve). The error bars represent $\pm \sqrt{k_i}$, where $k_i$ is the number of event in the bin of index $i$.}
\label{Fig:HistZ}
\end{figure}

The steps described above are combined to perform the full interferometric sequence depicted in Fig.~\ref{Fig:Schematics}b. Adjusting the phase shifter tilt $\epsilon/h$ to $-349$\,Hz, we record the final population imbalance $z = n/N_\text{t}$ as a function of the phase accumulation time $t_{\phi}$. For each value of $t_{\phi}$, we typically perform 18 independent measurements of $z$, and compute the ensemble average and standard deviation of $z(t_{\phi})$. As expected, we observe that $\langle z \rangle$ oscillates at the frequency $\epsilon/h$. Phase diffusion manifests as a damping of the oscillations associated with increasing fluctuations of $z$. Note that the error bars of Fig.~\ref{Fig:Schematics}c represent $\pm 1$ standard error of the mean. For $t_{\phi} \gtrsim 20$\,ms, $\langle z \rangle$ remains close to 0 because of the partial randomization of the phase while single experimental runs can still achieve $\sim 40\,\%$ imbalance.

To check the consistency of this data with that obtained by time-of-flight recombination (Fig.~\ref{Fig:PhaseShifter}), we derive the expression for $\langle z (t_{\phi}) \rangle $ at the output of the recombiner. We assume that the relative phase follows a normal distribution with time-dependent mean and variance given respectively by Eqs.~\eqref{Eq:PhaseShift} and \eqref{Eq:PhaseSpread}. Equation~\eqref{Eq:PhaseShift} describes the phase accumulation while Eq.~\eqref{Eq:PhaseSpread} accounts for phase diffusion. Assuming that the recombiner acts on the BECs as explained above (cf.\ Eq.~\eqref{Eq:contrast}), we can integrate over the phase distribution to obtain the expression for $\langle z \rangle$ at any time:
\begin{equation}
  \langle z(t_\phi) \rangle = C e^{-\Delta \phi_0^2/2}e^{-(R t_\phi)^2/2} \sin(\varphi_0+\epsilon \, t_\phi / \hbar) .
\label{Eq:PopDiff}
\end{equation}
The initial phase spread is responsible for a reduced contrast of the fringes at $t = 0$, while phase diffusion causes a Gaussian damping of the oscillations of $\langle z \rangle$. We set the phase diffusion rate $R$, the initial phase spread $\Delta \phi_0^2$ and the detuning $\epsilon$ to the values obtained from the interference pattern analysis (see Fig.~3). The contrast is self-consistently extracted from the data of Fig.~1c by binning all the single-shot population imbalances (see Fig.~\ref{Fig:HistZ}) and assuming that $\phi$ samples the interval $\left[0, 2\pi\right[$ uniformly. In the absence of noise, the probability density function of $z$ is given by
\begin{equation}
f(z) = \frac{1}{\pi}\frac{1}{\sqrt{1-\left(z/C\right)^2}} 
\label{Eq:pdfZ}
\end{equation}
for $|z|<C$, and $0$ elsewhere.

To account for technical noise of the Mach-Zehnder interferometer, Eq.~\eqref{Eq:pdfZ} is convolved with a Gaussian function of mean 0 and standard deviation $\sigma_z$~\cite{Geiger2011}. A fit to the data (red curve of Fig.~\ref{Fig:HistZ}) gives the values of $C = 0.42$ and $\sigma_z = 0.075$.

Equation \eqref{Eq:PopDiff} is used as a fit model for the mean population imbalance of Fig.~\ref{Fig:Schematics}c (solid red line) with only free parameters $\varphi_0$ and a constant offset to account for imperfect balancing of the double well. From the technical noise and the contrast extracted from the data of Fig.~\ref{Fig:HistZ}, we estimate the noise on the phase estimation of the non-adiabatic recombiner as $\delta \phi = \Delta n / | \partial n / \partial \phi |_{\phi=0} \simeq \sigma_z/C \simeq 0.18$\,rad.

The good agreement of the model with the measurement confirms that the interrogation time of our interferometer is limited exclusively by phase diffusion. The black line in Fig.~\ref{Fig:Schematics}c is the signal expected in the absence of number squeezing. It underlines that the interrogation time is extended by the use of a number-squeezed state.

\section{Discussion}

We can now compare the phase estimation performance of our non-adiabatic recombiner to that of the usual time-of-flight recombination scheme. The intrinsic read-out noise of the non-adiabatic recombiner $\delta \phi = 0.18$\,rad is about twice the detection noise of time-of-flight recombination $\delta \phi_\text{d} = 0.07$\,rad (see Methods). This difference is essentially due to the moderate contrast ($42\,\%$) of the non-adiabatic recombiner. Were it increased to $100\,\%$, for instance through optimal control of the double-well potential~\cite{Grond2009, Grond2009a, Grond2011},the sensitivity of the non-adiabatic recombiner would compete with that of the time-of-flight technique.

We also underline that phase estimation based on atom counting is more robust than phase extraction from interference patterns. Firstly it does not require high imaging resolution, secondly, using a recombiner to convert the relative phase into a population imbalance allows the use of the precise atom counting techniques already available~\cite{Esteve2008, Bucker2009, Lucke2011}. Contrary to fringe fitting, it does not rely on time of flight, which makes it more suitable for a fully integrated device on a chip~\cite{Heine2010}.

A more fundamental distinction between both methods is their potential sensitivity limits. Even though phase estimation based on a fit to the time-of-flight interference pattern can reach sub-shot-noise sensitivity, that is $\delta \phi_\text{fit} \leq \delta \phi_\text{SQL} \propto 1/\sqrt{N}$, it is fundamentally bounded by $N^{-2/3}$~\cite{Chwedenczuk2012}. This lower bound holds for any entangled state used in the interferometer. This non-trivial result underlines that the Heisenberg scaling of the phase sensitivity $\delta \phi \propto 1/N$ will not be accessible to atom interferometers using this phase estimation strategy. On the contrary, our non-adiabatic recombiner essentially works as a conventional optical beam splitter, and we thus conjecture that it can reach the Heisenberg scaling~\cite{Pezze2005, Grond2011}.

In conclusion, we have developed a new interferometry technique for trapped and spatially separated Bose-Einstein condensates based on a novel, non-adiabatic recombiner. The geometry is reminiscent of the familiar optical Mach-Zehnder interferometer. Because its two arms are spatially separated, our device is sensitive to differentially applied forces. The tight confinement and precise positioning on an atom chip allows for local probing. We have demonstrated that using a number-squeezed state helps increasing the coherence time, which in turn is still limited to $\sim 30$\,ms by dephasing due to interactions.

Our observations illustrate a fundamental limitation of interferometry with trapped Bose-Einstein condensates for precision measurements: interactions dominate the randomization of the phase. This is in contrast to previous experiments in which technical fluctuations of the energy difference or parasitic excitations during splitting were the dominant cause of dephasing~\cite{Shin2004, Schumm2005}. 

Strategies for creating optimal input states, for example by means of optimal control of the double well, will improve upon our present interferometer only marginally~\cite{Grond2010}. To go far beyond requires, in addition, phase estimation protocols based on Bayesian analysis of the full atom number distribution after the recombiner~\cite{Grond2011}. Developing these techniques could yield both a sub-shot-noise sensitivity and a long interrogation time, even in the presence of interactions.

The non-adiabatic recombiner can also be seen as a new tool to perform rotations of the pseudo-spin describing the many-body wave function~\cite{Pezze2005} around the $x$ axis of the Bloch sphere. Such a tool was missing for condensates trapped in double-well potentials. This may prove useful for the tomography of the many-body wave function, e.g.\ to investigate strongly entangled states such as squeezed or Schrödinger cat states.

\section{Methods}

\subsection{Preparation of the condensate}

To minimize breathing oscillations during splitting, we prepare the initial quasi-condensate in a rf-dressed trap. The rf amplitude is slightly below the splitting threshold~\cite{Lesanovsky2006}. The potential is harmonic along the longitudinal ($z$) and vertical ($y$) directions, with trapping frequencies \mbox{$\nu_z=12.4$ Hz} and \mbox{$\nu_y = 1.75$ kHz} respectively (see Fig.~\ref{Fig:Schematics}a for the orientation of the axes). In the horizontal, slightly anharmonic direction ($x$), simulations predict a spacing between the ground and first excited state of 1.02 kHz. By means of rf evaporative cooling, we prepare samples with less than $10\%$ relative number fluctuations from shot to shot. The temperature is estimated from the density profile after time of flight using a stochastic model for phase-fluctuating quasi-condensates~\cite{Stimming2010}.

\subsection{Radio-frequency dressing}

The rf signal is sent through two wires which are parallel to the main trapping wire~\cite{Trinker2008a} (cf.\ Fig~\ref{Fig:Schematics}a). In order to turn the trap into a double-well potential, the frequency of the rf field is kept constant $30$\,kHz below the atomic Larmor frequency of $910$\,kHz due to the static fields ($m_F = 0 \rightarrow m_F = \pm 1$ transitions), while the amplitudes in the two wires are ramped up to $52$\,mA peak to peak, creating an oscillating magnetic field of 1.7 G at the position of the atoms~\cite{Lesanovsky2006, Schumm2005}. The distance between the wells can be tuned by changing the total rf intensity. To tilt the double well, the intensities in the rf wires are varied independently from each other.

\subsection{Splitting}

The double-well potential obtained after splitting is characterized by the trap frequencies of each individual well $\nu_z = 13.2$\,Hz and \mbox{$\nu_y = 1.84 $\,kHz}. Along the splitting direction, we measure for each well \mbox{$\nu_x = 1.44 $\,kHz}. The two minima are separated by $2\,\mu$m and simulations predict a barrier height of $h\times3.7$\,kHz and a tunnel coupling energy of $h \times 0.1$\,Hz~\cite{Raghavan1999}.

\subsection{Calibration of the trap simulations}

We compute the geometry of the rf-dressed potential beyond the rotating-wave approximation by means of a Floquet analysis~\cite{Hofferberth2007}. The trap is first simulated from the knowledge of the chip design (see Fig.~\ref{Fig:Schematics}a) and the values of the control parameters. The latter are then adjusted in the simulation through independent measurements of the trap properties as follows: for the non-dressed trap, we measure precisely the current in the trap wire and the current in the wires responsible for the longitudinal confinement. This leaves two free parameters to fully define the static trap: the magnetic field $B_0$ at the centre of the trap and the bias field $B_x$, which are adjusted to match the measured atomic Larmor frequency and the measured transverse trap frequency. For the rf-dressed trap, the absolute amplitudes of the rf current in each dressing wire are calibrated independently by measuring the trap bottom with one wire at a time. Then both wires are switched on and the relative phase between the rf currents is fine-tuned for the trap bottom of the dressed trap to match the simulation for different values of the global rf intensity. The good agreement of the simulation with the experiment is checked by measuring the distance between the wells (from the fringe spacing of the interference patterns) and the trap frequencies. Furthermore, measurements of the frequency of Josephson oscillations in strongly coupled double wells confirm the good calibration of the trap simulation used for example to simulate the recombiner.

\subsection{Time-of-flight number and phase measurement}

At each point in the sequence, we can measure the relative phase and the atom number difference between the two condensates. To measure the phase, we abruptly switch off the trap, let the clouds overlap and image the interference pattern after $46$\,ms of free expansion using our light-sheet fluorescence detection system~\cite{Bucker2009}. We extract the phase from the Fourier transform of the transverse density profile after integration along the longitudinal axis. Since the atoms are projected onto the three Zeeman sub-levels of the $F = 1$ manifold at switch-off, and to avoid the smearing of the interference patterns inevitably caused by stray fields, a magnetic gradient is applied during time of flight to separate the clouds. The Fourier analysis is performed only on the atoms in $m_F = 0$ (more than half of the total atom number).

Two different methods were used to access the atom number difference $n = N_\text{L} - N_\text{R}$. For the data of Fig.~\ref{Fig:Fluctuations}a, the dressing intensity was suddenly raised to increase the splitting distance and give opposite momenta to both clouds. For the data of Fig.~\ref{Fig:Schematics}c, the rf dressing intensity was quickly ramped down to initiate a transverse inward motion in the static harmonic potential. After the atoms acquired a sufficient velocity, the trap was switched off and the two clouds crossed and spatially separated. In both cases, we ended up with two well-separated clouds after time of flight, from which we count the fluorescence photons in regions of equal size.

\subsection{Number distribution analysis}

To measure the atom number difference distribution, we performed the same analysis as in Ref.~\cite{Bucker2011}. Assuming that the number of fluorescence photons detected from each atom is a random variable of mean $p$ and variance $\sigma_p^2$, and including a random background of variance $\sigma_\text{b}^2$, the variance of the photon number difference between the two counting areas is
\begin{equation}
\Delta s^2 = p^2 \xi_\text{N}^2 N + N \sigma_p^2 + \sigma_\text{b}^2 ,
\label{Eq:FlucN}
\end{equation}
where $\xi_\text{N} = \Delta n/\sqrt{N}$ is the number-squeezing factor. For our detector, the mean of the photon number distribution is $p = 16 \pm 1$\,photons/atom. Photon shot noise and the amplification noise of the EMCCD camera yield $\sigma_p^2 = 2 p$. Using more than 200 fluorescence pictures to measure $\Delta s^2$ and $\sigma_\text{b}^2$, we get the value $\xi_N = 0.41 \pm 0.04$ (without correcting for imaging noise, we measure $\xi_{N,\text{unc}} = 0.52$). Setting $\xi_\text{N} = 0$ in Eq.~\eqref{Eq:FlucN} enables to estimate the detection noise (dotted red curve in Fig.~\ref{Fig:Fluctuations}a).

\subsection{Phase distribution analysis}

To compare the phase distribution measured on Fig.~\ref{Fig:Fluctuations}b with the one expected from a coherent state, we simulated the full phase extraction method. We first computed the one-body density distribution corresponding to the ballistic expansion of a symmetric superposition with $\phi=0$, and generated the positions of the atoms for each realization by drawing $N_{\text{tof}}$ independent atom positions from this distribution. Since each atom is drawn independently, this procedure mimics a coherent state. We used $N_{\text{tof}}=500$ to take into account the reduction of the number of atoms we extracted the phase from due to the Stern-Gerlach separation during time of flight. Taking into account the photon shot noise, the amplification noise of the CCD camera and the optical resolution of our imaging system (we modelled the optical point-spread function by a Gaussian with RMS width of $10\,\mu$m, to be compared to the fringe spacing of $107\,\mu$m), we produced a large number of interference patterns ($\sim 1,000$), on which we ran our Fourier phase extraction routine. We obtained a Gaussian phase distribution with standard deviation $\delta \phi_\text{tof} = 0.08$\,rad (dot-dash green curve of Fig.~\ref{Fig:Fluctuations}b), significantly larger than the shot noise limit expected for an ideal detection system $\delta \phi_\text{SQL} = 1/\sqrt{N} \simeq 0.03$ (dashed blue curve).

A rough estimate of the phase detection noise is given by $\delta \phi_\text{d} = \left(\delta \phi_\text{tof}^2 - \delta \phi_\text{SQL}^2\right)^{1/2} \simeq 0.07$\,rad, showing that detection noise dominates over projection noise.

\subsection{Test of uniform phase}

In Fig.~\ref{Fig:PhaseShifter}b we used a Rayleigh non-uniformity test~\cite{Fisher1995} to distinguish the points having a uniform phase distribution (black points) from the others (blue points). The blue points correspond to distributions whose p-values were below $0.05$, indicating a probability less than $5\%$ to be compatible with a uniform distribution. Only those points were fitted with Eq.~\eqref{Eq:PhaseSpread}.

\subsection{Uncertainties and error bars}

The uncertainties given in the text, as in $\Delta \phi = 0.16 \pm 0.01$\,rad or $\xi_\text{N}^2 = -7.8\pm0.8$\,dB are statistical uncertainties calculated as follows: for directly measured quantities they are the standard error of the mean assuming Gaussian distributions, for fitted parameters we use the standard error obtained from the fit, and for calculated values we use usual error propagation methods.

\subsection{Acknowledgments}
\begin{acknowledgments}
We are grateful to Thomas Betz, Stephanie Manz and Aurélien Perrin for earlier work on the experiment. We thank Igor Mazets, Remi Geiger, Jan Chwede\'{n}czuk and Patrizia Vignolo for helpful discussions. JFS acknowledges the support of the Austrian Science Fund (FWF) through his Lise Meitner fellowship (M 1454-N27). TB and RB acknowledge the support of the Vienna Doctoral Program on Complex Quantum Systems (CoQuS). This research was supported by the European STREP project QIBEC (284584), the European Integrated project AQUTE (247687) and the FWF projects SFB FoQuS (SFB F40) and CAP (I607-N16).
\end{acknowledgments}

\subsection{Author contributions}
TB and JFS performed the experiments and analysed the data. SvF, TB, and JFS performed numerical simulations. All the authors contributed to the elaboration of the experiment and to the final manuscript.

\clearpage

\renewcommand{\figurename}{Supplementary Figure}
\renewcommand{\thefigure}{S\arabic{figure}}
\renewcommand{\theequation}{S\arabic{equation}}
\renewcommand{\refname}{Supplementary References}

\section{SUPPLEMENTARY INFORMATION}

\section*{Supplementary Notes}

\subsection*{Supplementary Note 1: decoupling and phase diffusion} From the trap calibration (see Methods) we estimate a tunnel coupling energy~[53] in the symmetric double well just after splitting of the order of $h \times 0.1$\,Hz, high enough to prevent phase diffusion. However, when we tilt the double well even by a few degrees, phase diffusion is observed. Since the currents in the dressing wires are varied so as to keep the distance between the minima constant, the coupling energy is not expected to vary much when the double well is tilted. Nevertheless, we checked that the phase diffusion rate we measured in the tilted trap was compatible with the one observed when splitting a condensate even more into a symmetric double well with vanishing tunnel coupling energy ($< h \times 10^{-8}$\,Hz). This is consistent with simulations of the phase dynamics confirming that even for the lowest tilt angles we used, the energy detuning between the wells was strong enough with respect to the tunnel coupling to effectively decouple the condensates and enable phase diffusion.

\subsection*{Supplementary Note 2: Estimation of the chemical potential and phase diffusion rate after splitting}

\begin{figure}[h]
\centering
\includegraphics[width=\linewidth]{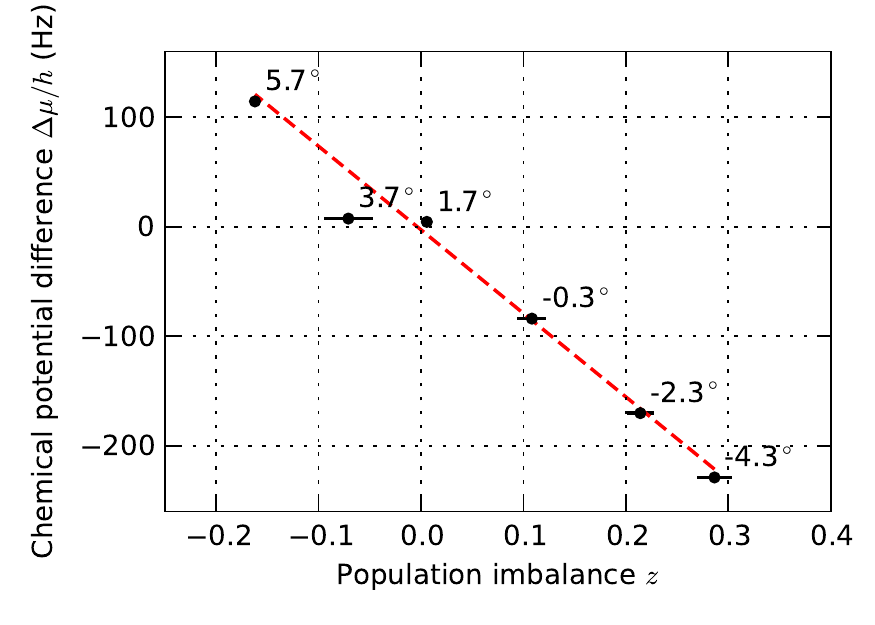}
\caption{\textbf{Direct measurement of $\Delta \mu(z)$ when the condensate is split to yield a population imbalance.} The numbers labelling the points are the angles (in $^\circ$) used during splitting to obtain the measured imbalance. The obtained population imbalance is responsible for $\Delta \mu$ measured in the levelled double well. The dashed line is a linear fit to the data from which $g n(0)$ is extracted according to Eq.~\eqref{eq:Dmu}. The horizontal error bars are $\pm 1$ standard error of the mean. The vertical error bars representing $\pm 1$ standard error of the fitted energy difference are smaller than the points.
\label{fig:Dmu_vs_z}}
\end{figure}

Because the splitting is fast compared to the period $T_\text{b}$ of the axial breathing mode ($5\,\text{ms} \ll T_\text{b} = (\sqrt{3} \, \nu_z)^{-1} \simeq 44$\,ms), the process is not adiabatic for the axial motion but, on the contrary, can be considered sudden. Right after splitting, the Thomas-Fermi (TF) radius is lager than its new equilibrium value because each BEC sees its density suddenly divided by 2. Although this effect is partially compensated by a slight increase of $\nu_z$ and a change of the 1D interaction constant during splitting, the axial breathing mode is excited and the condensates are not at equilibrium. Consequently, their chemical potentials (e.g.\ entering Eq.~(3)) must be replaced by the ``effective'' chemical potentials $\mu_i^\text{eff} = \hbar \, \mathrm{d} \phi_i/\mathrm{d} t$, where $\phi_i(t)$ is the time-dependent global phase of each BEC's wave function subject to a breathing mode ($i\in\{R,L\}$, $t=0$ here corresponds to the end of splitting). With our parameters, the BECs are well within the 1D TF regime. Transversally, they are essentially in the ground state of the single-particle Hamiltonian (negligible interactions). The linear density profiles of the breathing BECs are well described by 1D TF profiles whose radii $R_i(t) = b_i(t) R_i^\text{eq}$ evolve according to the equation~\cite{Kagan1996}
\begin{equation}
\ddot{b}_i + \omega_z^2 \, b_i = \omega_z^2/b_i^2 .
\end{equation}
Here, $R_i^\text{eq}$ denotes the equilibrium TF radii corresponding to each BEC. The corresponding phases do not evolve linearly with time but read
\begin{equation}
\phi_i(t) = \mu_i^\text{eq} \tau_i(t) / \hbar + \epsilon_i t / \hbar , 
\end{equation}
where $\tau_i = \int_0^t b_i^{-1}(u) \, \mathrm{d}u$, the $\mu_i^\text{eq}$ are the chemical potentials that the condensates would have if they were at equilibrium, and the $\epsilon_i$ denote the potential energies at the center of the two wells. For short times, the phases can be approximated to first order in $t/T_\text{b}$. Assuming $\dot b_i(0) = 0$, as expected for a sudden splitting we obtain $\phi_i(t) \simeq \mu_i^\text{eq} t/(\hbar \, b_i(t=0))$, which yields
\begin{equation}
\mu_i^\text{eff} = g \, n_i(0) + \epsilon_i ,
\label{eq:effective_mu}
\end{equation}
$n_i(0)$ being the peak linear densities after splitting, and $g$ the 1D interaction constant after splitting. This expression has a simple interpretation: one recognizes the usual expression of the chemical potential of an equilibrium quasi-BEC in the 1D TF regime in which the equilibrium peak density has been replaced by the peak density right after splitting (one usually sets $\epsilon_i = 0$). 

\textit{Symmetric splitting.} Note that in the case of a symmetric splitting we have $n_\text{L}(0) = n_\text{R}(0) = n(0)/2$, where $n(0)$ is the peak density before splitting and the two clouds breath in phase and with the same amplitude. The relative phase is simply given by Eq.~(1) in which $\epsilon = \epsilon_\text{R} - \epsilon_\text{L}$. 

\textit{Imbalanced splitting.} We now turn to the estimation of the phase diffusion rate. We can write Eq.~\eqref{eq:effective_mu} as $\mu_i^\text{eff} = g N_i n(0)/N$ because sudden splitting leads to identical density profiles before and after splitting, even if there is an imbalance ($N_i \neq N/2$). We readily obtain
\begin{equation}
\left.\frac{\partial \mu}{\partial \mathcal{N}}\right|_{\mathcal{N} = N/2} = g \frac{n(0)}{N} ,
\end{equation}
which we use to evaluate the phase diffusion rate in Eq.~(3). From the knowledge of the trap frequencies, number of atoms, squeezing factor, and using a 1D TF theory we get $R = 46 \pm 4$\,mrad/ms.

Another quantity of interest is the difference of chemical potential $\Delta \mu = \mu_\text{L}^\text{eff} - \mu_\text{R}^\text{eff}$ which arises from a population imbalance $z \neq 0$. From Eq.~\eqref{eq:effective_mu} we obtain
\begin{equation}
\Delta \mu = g n(0) z = N z \left.\frac{\partial \mu}{\partial \mathcal{N}}\right|_{\mathcal{N} = N/2} .
\label{eq:Dmu}
\end{equation}
This enabled us to directly measure $\partial \mu/\partial \mathcal{N}$. For this, the condensate was split in a tilted trap ($\epsilon \neq 0$) and then levelled back to the horizontal ($\epsilon = 0$). Adjusting the tilt angle for splitting gave control over the population imbalance which then drove the phase evolution in the levelled double well (see Supplementary Figure~\ref{fig:Dmu_vs_z}). We measured this effect for 6 values of $z$ and indeed observed a linear dependence of $\Delta \mu$ with $z$ and found the slope $g n(0)/h = 763 \pm 53$\,Hz. This gives a third value of $R = 57 \pm 5$\,mrad/ms, in agreement with both the expected theoretical value mentioned above and the one directly measured (see main text).\\

\subsection*{Supplementary Note 3: Action of the recombiner} Here we derive Eq.~(4) of the main text. In the transverse dimensions of our quasi-condensates~\cite{Salasnich2002} interactions can safely be neglected for the transverse spatial dynamics, the non-linear term of the Gross-Pitaevskii equation being negligible compared the the kinetic term. We assume that the trapping potential stays symmetric with respect to a vertical symmetry plane, $V(-x,y,z,t) = V(x,y,z,t)$, which is expected from the atom chip symmetry. Before recombination, the wave function describing the two condensates has the same symmetry, except for a relative phase: $\psi_\text{BEC}(x,y,z,t_\text{BS}=0) = \left(\psi_\text{L} + \psi_\text{R}\right)/\sqrt{2}$, with $e^{i\phi} \psi_\text{R}(-x,y,z) = \psi_\text{L}(x,y,z) \equiv \psi(x,y,z,t_\text{BS}=0)$. From the symmetry of the potential and propagating with the Schrödinger equation, one finds that the BEC wave function can be written at all time as
\begin{equation}
\psi_\text{BEC} = \left[\psi(x,y,z,t) + e^{i\phi} \psi(-x,y,z,t)\right]/\sqrt{2} .
\label{Eq:WFRecomb}
\end{equation}
The imbalance $z \equiv \int_{x<0}|\psi_\text{BEC}|^2 \, \mathrm{d}\mathbf{r} - \int_{x>0}|\psi_\text{BEC}|^2 \, \mathrm{d}\mathbf{r}$ can readily be calculated from Eq.~\eqref{Eq:WFRecomb}, which gives Eq.~(4) of the main text. The contrast is essentially given by the overlap of $\psi(x,y,z,t)$ and $\psi(-x,y,z,t)$.\\

\subsection*{Supplementary Note 4: Simulations of the non-adiabatic recombiner}

\begin{figure}[h]
\centering
\includegraphics[width=\linewidth]{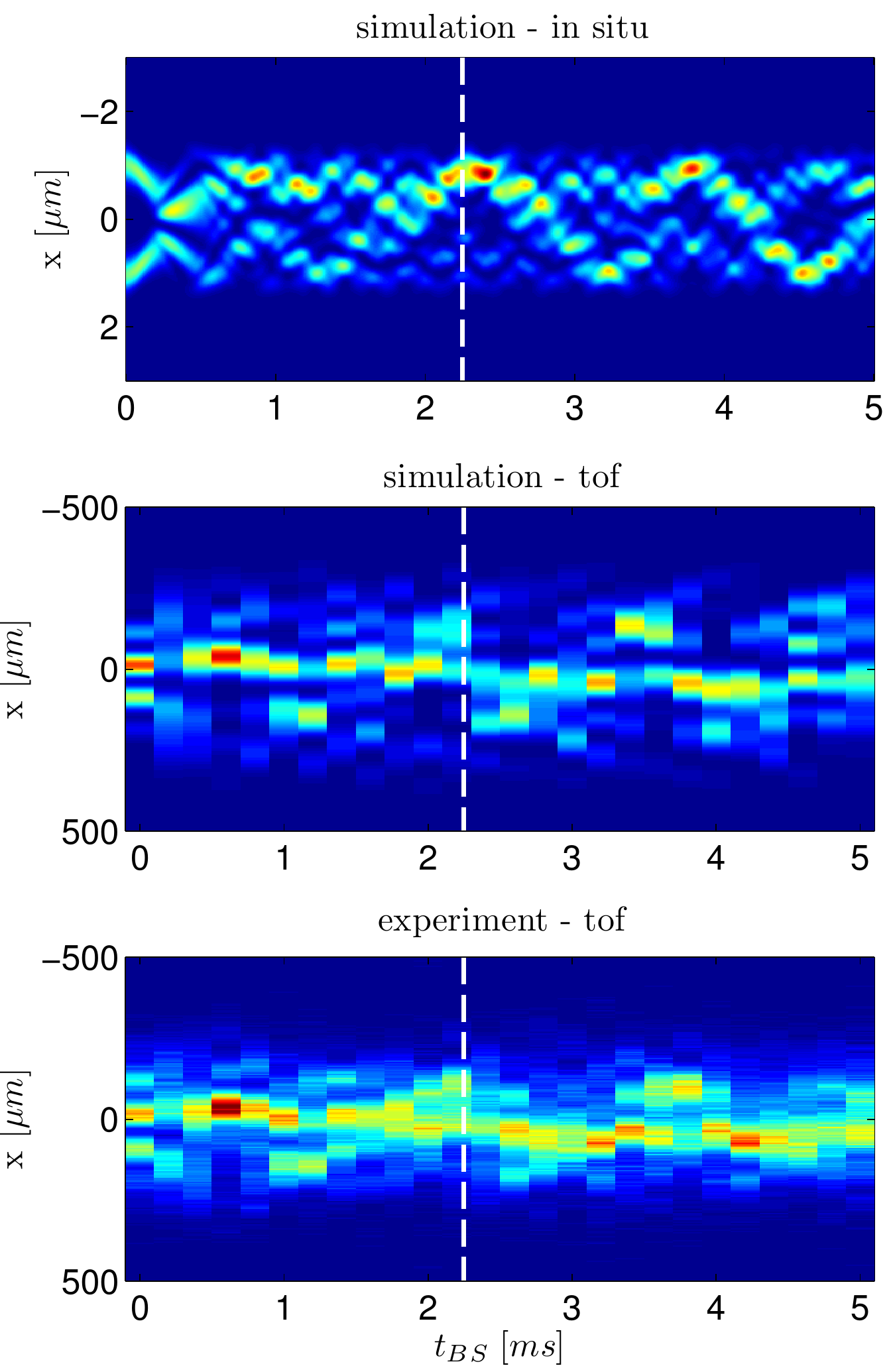}
\caption{\textbf{Dynamics of the condensates in the recombiner.} Using the phase-shifter, a state with a phase of $2.13$\,rad is prepared. Its dynamics in the recombiner is monitored by time-of-flight imaging. From top to bottom: simulation of the transverse \textit{in situ} density profile;  simulation of the momentum distribution scaled to the 46ms time of flight, including the finite imaging resolution; measured density profiles after expansion. The white dashed line corresponds to $t_{BS} = 2.25$\,ms giving maximum contrast. We checked the agreement between data and simulation for 8 initial values of the phase.}
\label{Fig:BSsim}
\end{figure}

We performed numerical simulations of the dynamics of the condensates during the non-adiabatic recombination using a 1D Gross-Pitaevskii solver in the double-well potential. The effective 1D interaction constant was computed assuming a quasi-condensate profile longitudinally~\cite{Gerbier2004} and a Gaussian profile corresponding to the non-interacting ground state in the vertical transverse direction (see Fig.~4a--b). The simulations show complex patterns arising from the multiple reflections and transmissions in the double-well potential.

To validate these simulations, we used the phase shifter to prepare states with 8 different relative phases and abruptly reduced the splitting similarly to the non-adiabatic recombination. After holding the atoms for an adjustable time $t_\text{BS}$, we eventually switched off the potential instead of separating the clouds. Since for our parameters, the transverse density distribution after time of flight is almost homothetic to the initial transverse momentum distribution, this gives us access to the momentum distribution at any time in the recombiner. We could compare it to the results of the simulation and found qualitative agreement for all phases (see Supplementary Figure~\ref{Fig:BSsim}).

\subsection*{Supplementary Note 5: 1D effects} It may seem surprising that phase fluctuations typical for quasi-condensates~[41] do not play a significant role here. Nevertheless we stress that the relevant quantity for interferometry is the relative phase between the condensates and not the spatially fluctuating phase of a single BEC. Given our atom number, squeezing factor, and temperature, the relative phase correlation length~\cite{Kitagawa2011} exceeds the length of the condensates. Experimentally, this is confirmed by the observation of a high contrast of the interference patterns integrated over the whole axial length of the cloud more than $80$\,ms after splitting (above $60\,\%$). It therefore makes sense to define a single global relative phase between the two quasi-condensates and to neglect 1D phase fluctuations. For longer times, however, the contrast of individual interference patterns starts to decrease, suggesting that 1D effects might be an issue when trying to enhance further the interrogation time. This could be readily solved by using true condensates instead of quasi-condensates.

\subsection*{Supplementary Note 6: Contributions to the energy difference $\epsilon$} The dominant contribution to the energy $\epsilon$ is the difference of gravitational potential  between the two trap minima. A difference of $100$\,Hz corresponds to a vertical shift of $47$\,nm, which is just a fraction of the BEC's transverse width. Nevertheless, two other effects may contribute to $\epsilon$: different values of the magnetic field between the two minima, investigated in Ref.~[22], and the variation of the trapping frequencies during the tilt, which shifts both the zero-point energies and the chemical potentials of the BECs. From simulations of the rf-dressed trap we estimate these two contributions to be below $10\%$ and $2\%$ respectively. Uncontrolled inhomogeneous electric fields close to the chip could also contribute to $\epsilon$. For these reasons, making a precision measurement of gravity seems challenging.

\end{document}